\documentclass[9pt,twocolumn,twoside]{article}
\usepackage{amsmath}
\usepackage{hyperref} % For hyperlinks in the PDF

\usepackage[english]{babel}
\usepackage[normalem]{ulem}

\usepackage{cleveref}

\usepackage[usenames, dvipsnames]{color}

\usepackage{titling} % Customizing the title section

\usepackage{graphicx}
\usepackage{cite}%

\usepackage{authblk}
\usepackage{textcomp}
\title{Inline self-diffraction dispersion-scan of over octave-spanning pulses in the single-cycle regime} 

\author[1,*]{Miguel Canhota}
\author[1,2]{Francisco Silva}
\author[3]{Rosa Weigand}
\author[1,4]{Helder M. Crespo}

\affil[1]{IFIMUP-IN and Departamento de F\'isica e Astronomia, Faculdade de Ci\^encias, Universidade do Porto, R. do Campo Alegre 687, 4169-007 Porto, Portugal.}
\affil[2]{Sphere Ultrafast Photonics, Parque de Ci\^{e}ncia e Tecnologia da Universidade do Porto, R. do Campo Alegre 1021, Ed. FC6, 4169-007 Porto, Portugal.}
\affil[3]{Departamento de \'{O}ptica, Facultad de Ciencias F\'{i}sicas, Avda. Complutense s/n, Universidad Complutense de Madrid, 28040, Madrid, Spain.}
\affil[4]{Grupo de Investigaci\'{o}n en Aplicaciones del L\'{a}ser y Fot\'{o}nica (ALF-USAL), Universidad de Salamanca, E-37008, Salamanca, Spain.}
\affil[*]{Corresponding author: mcanhota@fc.up.pt}

\date{Compiled \today}

\begin{document}

\maketitle

\begin{abstract}
We present an implementation of dispersion-scan based on self-diffraction (SD d-scan) and apply it to the measurement of over octave-spanning sub-4-fs pulses. The results are compared with second-harmonic generation (SHG) d-scan. The efficiency of the SD process is derived theoretically and compared with the spectral response retrieved by the d-scan algorithm. The new SD d-scan has a robust inline setup and enables measuring pulses with over-octave spectra, single-cycle durations and wavelength ranges beyond those of SHG crystals, such as the ultraviolet and the deep-ultraviolet.

\end{abstract}

\section{Introduction}
Intense single-cycle light pulses are an important tool for attosecond science and high-field physics, and are now enabling a new generation of laser-plasma accelerators \cite{Guenot2016}. These pulses can be generated by several post-compression techniques \cite{Wirth2011, Silva2014a, Balciunas2015} where precise temporal characterization is crucial for optimizing their compression and for establishing the pulse duration in an experiment. For example, a chirp change of merely 4\,fs$^{2}$ in the driving pulse resulted in a 20\% change in the accelerated electron energy \cite{Guenot2016}.   Sub-cycle 0.975\,fs optical pulses have been measured with attosecond streaking \cite{Hassan2016}, and a petahertz optical oscilloscope was demonstrated with two-cycle pulses \cite{Kim2013}, but these strong-field techniques demand high pulse energies, vacuum beamlines, isolated attosecond pulses, and electron or extreme ultraviolet spectral detection.

The optical measurement of single-cycle pulses usually requires nonlinear media with low dispersion and large phase-matching bandwidths, due to the over octave-spanning spectra and extremely short durations of the pulses. In the case  of frequency-resolved optical gating (FROG) \cite{trebino2012frequency}, these characteristics translate into an overlap between fundamental and SHG spectra that limits the measurement bandwidth to less than one octave. This can be overcome with noncollinear setups, but the resulting geometrical time smearing limits the temporal resolution and gives overestimated pulse lengths. Addressing these problems has required precise knowledge of the total spectral response function affecting the measured FROG signal, which includes effects due to the noncollinear geometry, nonlinear crystal thickness and phase matching bandwidth (all carefully chosen for a particular pulse), dispersion of the nonlinearity and detector sensitivity \cite{Baltuska1999}.
%This approach enabled FROG measurements of 1.7-cycle pulses at 790\,nm (4.5\,fs) \cite{Baltuska1999}.
Noncollinear cross-correlation FROG (XFROG) %provides added control over phase-matching and
was recently used to measure %synthesized
0.9-cycle, 4.2\,{$\mu$}m pulses (12.4\,fs) \cite{Liang2016}, but this required a fully characterized short reference pulse and the geometric time smearing is no longer negligible for few-fs pulses.
A variant of spectral phase interferometry for direct electric-field reconstruction (SPIDER) \cite{Iaconis1998}, spatially encoded arrangement (SEA)-SPIDER \cite{Wyatt2006}, is free of time smearing and enabled measuring 0.9-cycle pulses at 1.6\,{$\mu$}m (4.5\,fs) %, whereas FROG was limited by smearing to 8\,fs
\cite{Balciunas2015}.
%SEA-SPIDER encodes the information of the wavelength-dependent group delay into fringes in the spatial domain, hence greatly relaxing the stringent delay calibration and stability requirements of standard SPIDER. %
%SEA-SPIDER is free of time smearing, but the experimental setup is nevertheless complex, involving multiple noncollinear beams %with different frequencies and durations, and requiring careful spectral and spatial calibrations.%
Another SPIDER variant, two-dimensional spectral shearing-interferometry (2DSI) \cite{Birge2006,Birge2010}, %which generates fringes in the time domain and
was used to measure 1.1-cycle, 1\,{$\mu$}m pulses (3.7\,fs) \cite{Cox2012}. Very recently, time-domain ptychography was applied to 3.7\,fs, 1.5-cycle pulses at 800\,nm \cite{Witting2016}.
%Other optical pulse characterization methods have been developed that can access the few-cycle regime, %such as self-referenced spectral interferometry \cite{Trabattoni2015},
%but to our knowledge they have not yet been demonstrated with single-cycle pulses.

The single-cycle-capable optical techniques described above can be powerful but involve operations such as temporal overlap of short pulses, beam splitting and recombination, which all add to increased experimental complexity and difficulty.
The dispersion-scan (d-scan) technique \cite{Miranda2012} is a recent approach for the simultaneous measurement and compression of femtosecond laser pulses and was originally proposed as a way to simplify such steps, by coupling a compressor with a single-beam, non-interferometric, nonlinear measurement stage.
%resulting in a straightforward pulse measurement setup while still providing robust and precise pulse retrieval.
Experimentally, it involves measuring the spectrum of a nonlinear process, such as SHG, as a function of compressor position around
%, such as prism, grating, grism or chirped mirror compressors.
the maximum compression point (the reference position), where the dispersion introduced by each displacement step of the compressor is either well-known \cite{Miranda2012} or self-calibrated from the measurement \cite{Alonso2017}.
%In SHG d-scan, a %2D trace of the SHG spectrum vs. dispersion, from which the spectral phase of the pulse can be retrieved.
%two-dimensional trace of the SHG spectrum vs. position is obtained from which\label{eq:pulse_frequency_domain}
If we consider a pulse in the frequency domain,
\begin{equation}
\tilde{E}(\omega)=\left|\tilde{E}(\omega)\right|e^{i\phi(\omega)},
\label{eq:pulse_frequency_domain}
\end{equation}
where $\tilde{E}$ is the pulse electric field and $\phi$ is its spectral phase, the measured d-scan trace, $S_{\mathrm{meas}}$, can be written as the product of a spectral response function, $R(\omega)$, and an ideal trace, $S_{\textrm{ideal}}$ \cite{Miranda2012}
\begin{equation}
S_{\mathrm{meas}}=R(\omega) \times S_{\mathrm{ideal}} \equiv R(\omega) \left|\int_{-\infty}^{+\infty}E_{\mathrm{NL}}e^{-i\omega t}\mathrm{d}t\right|^{2},
\label{eq:SHG_d-scan}
\end{equation}
where $E_{\mathrm{NL}}$ is the dispersion-dependent nonlinear signal, which for SHG d-scan is the square of the time-domain field, $E(t,\zeta)$, after the compressor, i.e.
\begin{equation}
E_{\mathrm{NL}}= E^{2}(t,\zeta) \propto \left(\int_{-\infty}^{+\infty}\tilde{E}(\Omega)e^{-i\beta(\Omega)\zeta}e^{i\Omega t}\mathrm{d}\Omega\right)^{2},
\label{eq:E_field_SHG}
\end{equation}
with $\beta(\Omega)$ the frequency-dependent phase per unit displacement introduced by the compressor and $\zeta$ the compressor position. An optimization algorithm is then used to retrieve {\sl both} the spectral phase of the pulse, $\phi(\omega)$, and the unknown response function, $R(\omega)$, from the d-scan trace and calibrated linear spectrum \cite{Miranda2012}. A recent approach to fast d-scan retrieval enables obtaining the pulse amplitude and phase from the d-scan trace alone, but in this case the trace itself must be calibrated \cite{Miranda2017}.

SHG d-scan has been successfully demonstrated with few-cycle pulses since its inception \cite{Miranda2012,Miranda2012b} and is nowadays an established technique in the demanding sub-4-fs regime (see, e.g., \cite{Chang2016}) but a common misconception is assuming that SHG d-scan is limited to sub-octave pulses, because it uses collinear \cite{Miranda2012} or near-collinear \cite{Miranda2012b, Silva2014a} SHG. On the contrary, over octave-spanning single-cycle pulses have been measured directly with SHG d-scan, both in scanning \cite{Silva2014a,Miranda2017,Guenot2016} and single-shot \cite{Fabris2015} configurations. The fact that broadband SHG corresponds to sum-frequency generation (SFG) between all the frequencies in the spectrum translates into intrinsic redundancy in the d-scan trace: phase information of a particular spectral region of the pulse is found not only at its SHG frequency, but convoluted across the trace. This relaxes phase-matching and measurement bandwidth requirements, as a partial measurement of the SHG/SFG trace is sufficient for phase retrieval over the whole spectrum \cite{Miranda2012,Miranda2012b,Silva2014a,Fabris2015}. Few-cycle capable third-harmonic generation (THG) d-scan variants have also been devised \cite{Silva2013,Hoffmann2013}, but both SHG and THG processes are limited by transparency and phase-matching. This is particularly problematic for ultraviolet pulses, where choice of adequate SHG crystals is very limited. Also, the SHG or THG signal can easily extend into the deep- and vacuum-ultraviolet, further complicating its detection. On the other hand, degenerate third-order processes such as cross-polarized wave (XPW) generation and self-diffraction (SD) can offer greater flexibility, as they facilitate phase-matching over wide bandwidths, are practically wavelength-independent over the transparency zone of the nonlinear medium, and the signal lies in the same spectral region as the pulse to be measured.
%The development of SD and other 3rd-order optical diagnostics is therefore timely and relevant for the community.
SD has been applied to FROG \cite{trebino2012frequency} and very recently to SPIDER \cite{Birkholz2015}. XPW d-scan was recently demonstrated with sub-octave 6.5\, fs pulses \cite{Tajalli2016}, which required increasing their degree of linear polarization using Brewster reflections prior to measurement.

In this Letter we introduce a new inline d-scan scheme based on self-diffraction in thin transparent media (SD d-scan) and demonstrate it with over octave-spanning sub-4-fs pulses, further illustrating its potential for pulse measurement over nearly 3 octaves. Furthermore, the SD process makes this technique suitable for pulses with arbitrary polarization and wavelength.

\section{Experimental}

\begin{figure}[h]
\centering
\includegraphics[width=1.0 \linewidth]{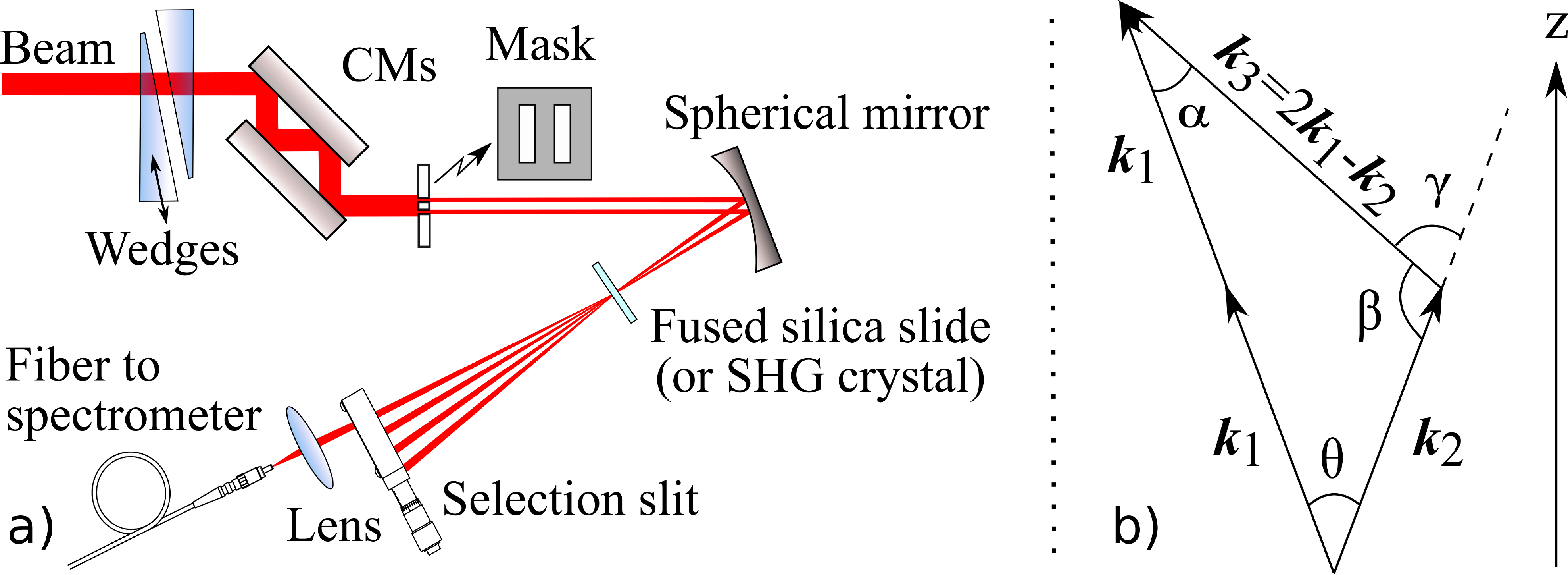}
\caption{a) Experimental setup for SD d-scan (see text for details). b) Wavevector diagram of noncollinear SD.}
\label{fig:setup}
\end{figure}

The experimental setup for SD d-scan (Fig.\,\ref{fig:setup}a) is analogous to SHG d-scan \cite{Miranda2012b,Silva2014a} and only requires replacing the SHG crystal (5\,{$\mu$}m BBO) with a thin (30\,{$\mu$}m) fused silica slide, whose dispersive pulse broadening is negligible even for single-cycle pulses.
%The dispersive pulse broadening experienced in the slide by a 3\,fs single-cycle Gaussian pulse at 800\,nm is 0.13\,fs, which is negligible.
The pulses are generated by a hollow-core fiber (HCF) compressor delivering sub-4-fs 800\,nm pulses with energy up to 200\,{$\mu$}J at 1\,kHz \cite{Silva2014a}, which includes the glass wedge and chirped mirror (CM) compressor shown in Fig.\,\ref{fig:setup}. The beam ($\approx 20$\,mm diameter) is sent through a mask with two vertical slits (2\,mm width and separation), placed so that spectra transmitted by each slit are identical, and also identical to the full-beam spectrum, although the latter is not a necessary requirement since differences with respect to the full spectrum can in principle be accounted for by the retrieved d-scan response function. The two pulses ($<1$\,{$\mu$}J) are focused in the slide (or in the BBO crystal) with a spherical mirror ($f = 25$\,cm) at a crossing angle of 1.4$^{\circ}$, and a movable slit is then used to select the off-axis SD beam (or the on-axis SHG beam) prior to the spectral measurement. We also measured the spectrum of the fundamental pulse after the fused silica slide for optimum pulse compression conditions (hence for maximum SD signal intensity) and no spectral changes were detected, which is compatible with the assumption that no significant self-phase modulation is taking place in the slide.
% and the introduction of a slitted mask over the central portion of the beam to split it into the two replicas necessary for SD,
%

The expression for the SD d-scan trace is also given by \eqref{eq:SHG_d-scan}, but now the nonlinear signal, $E_{\mathrm{NL}}$, is given by % \eqref{eq:SD/TG_d-scan}.
\begin{equation}
E_{\mathrm{NL}}(t,\zeta)=E^{2}(t,\zeta) E^{*}(t,\zeta)=\left|E(t,\zeta)\right|^{2} E(t,\zeta).
\label{eq:SD/TG_d-scan}
\end{equation}
The phase change introduced by the compressor
%, $\beta(\Omega)\zeta$,
is that of the moving wedges alone, i.e. $\beta(\Omega)\zeta=k(\Omega)l \equiv [n(\Omega)\Omega/c]l$, with $n$ the refractive index,
%of the wedge material,
$c$ the speed of light and $l$ the (relative) thickness of wedge glass crossed by the pulse.

\begin{figure}
\centering
\includegraphics[width=1.0\linewidth]{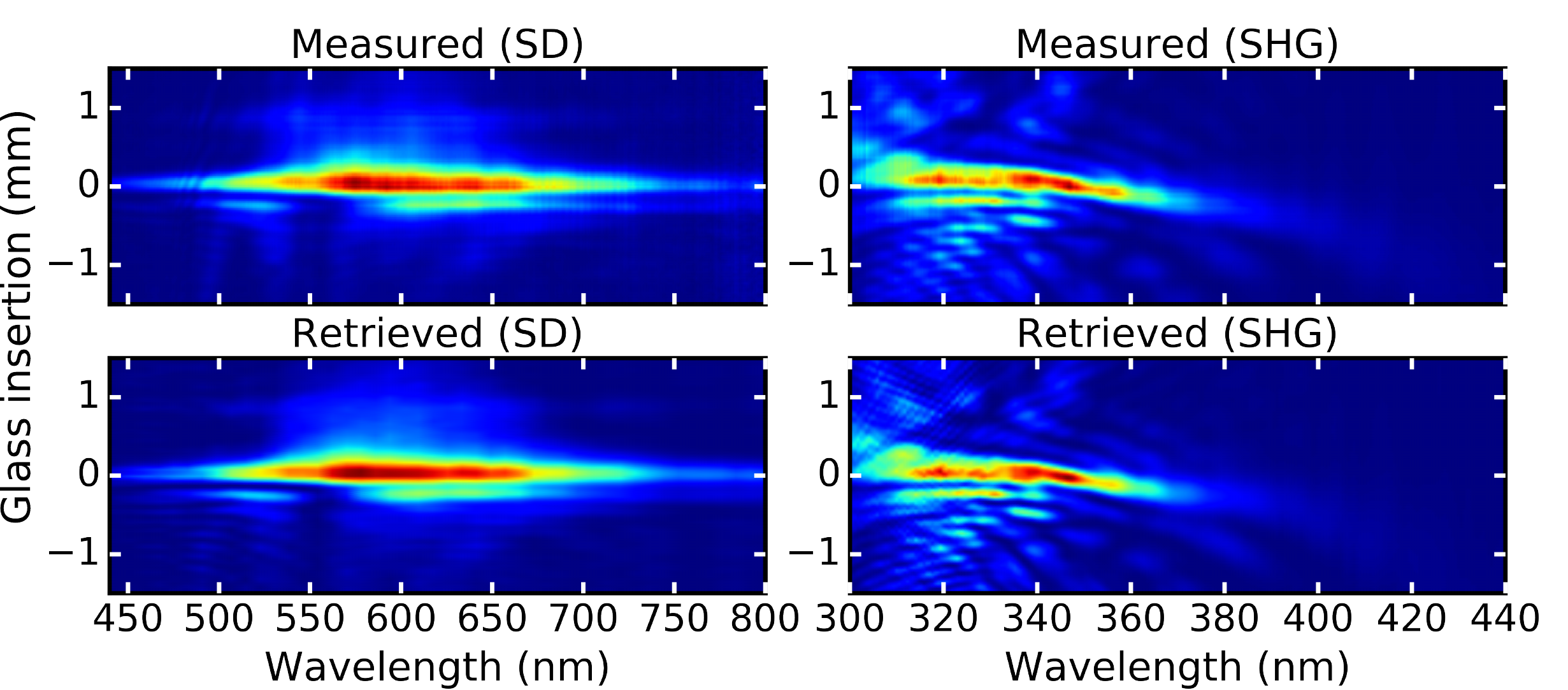}
\caption{Measured and retrieved SD and SHG d-scan traces.}
\label{fig:traces}
\end{figure}

The measured SHG and SD d-scan traces, each composed of 103 individual spectra (step size of 29\,{$\mu$}m), are given in Fig.\,\ref{fig:traces}, where a good visual agreement with the corresponding retrievals is observed.
The SD d-scan traces are smoother and less structured than their SHG counterparts due to the lower spectral phase sensitivity of SD (a 3rd-order process) compared to SHG.
The tilt in the SHG d-scan trace reveals some residual negative third-order dispersion (TOD) that was left uncompensated for (unlike in previous work \cite{Silva2014a}, where further propagation in a thin water cell resulted in single-cycle 3.2\,fs pulses). This enabled testing the SD d-scan with an over octave-spanning spectrum while retaining a more interesting temporal pulse structure due to TOD, hence showing that the method is sensitive to such a phase and may be used for its diagnostic and further correction.
%, not only in the visible to near-IR but also in harder spectral regions such as the ultraviolet.
% Both methods show a good visual agreement between measurement and retrieval. There is a total of 210 spectra. Step size is 29um.
%

\begin{figure}[h]
\centering
\includegraphics[width=1.0\linewidth]{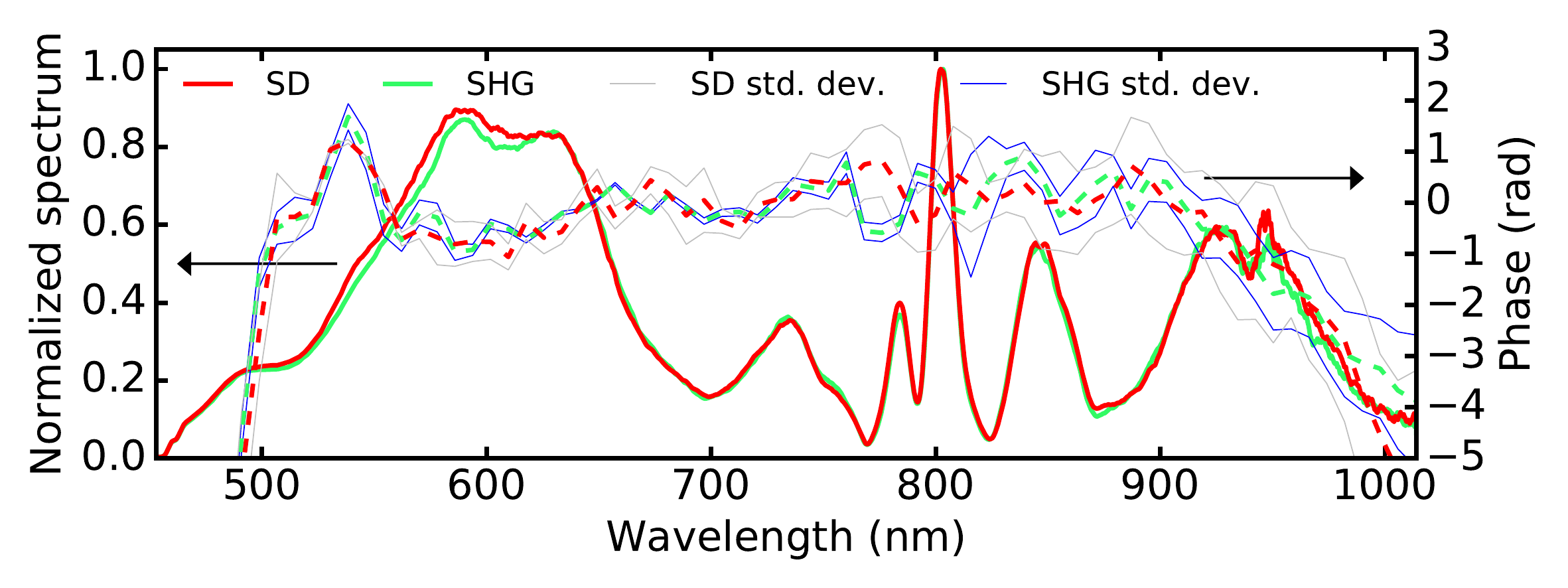}
\caption{Spectra and retrieved phases for SD and SHG d-scan.}
\label{fig:spectra_plots}
\end{figure}

Since SD is a degenerate process, i.e. $\omega_{\mathrm{NL}}=2 \times \omega - \omega=\omega$, one could at first expect the SD signal to cover the whole spectral range of the input pulse (see Fig.\,\ref{fig:spectra_plots}).
Instead, the SD d-scan trace only extends from $450$ to $800$\,nm (Fig.\,\ref{fig:traces}), since the efficiency of this process is %angle- and
wavelength-dependent, as shown in Appendix A and in Fig.\,\ref{fig:theoretical curves}a. This is not detrimental for retrieval over the full spectral range of the pulses, provided their phase does not change during propagation in the nonlinear medium (due to dispersion, nonlinear effects like self-phase modulation, or both), as required by \eqref{eq:SHG_d-scan}-\eqref{eq:SD/TG_d-scan}.
% otherwise a full propagation approach may be necessary.
Figure\,\ref{fig:spectra_plots} shows the measured spectra and retrieved spectral phases for SD and SHG d-scan, obtained by averaging 10 retrievals with random seed phases. For SD d-scan, we used 10 retrievals within one standard deviation of a set of 30, in order to minimize the contribution of phases affected by divergence problems in phase unwrapping due to the lower signal-to-noise ratio of SD at the edges.

Figure\,\ref{fig:spectra_plots} shows the measured spectra and retrieved spectral phases (described as 256-point vectors from 380 to 2400\,nm) for SD and SHG d-scan, obtained by averaging 10 retrievals performed with random seed phases.

The phases show a good agreement, within their standard deviations, diverging rapidly below $\approx 500$\,nm and rolling off after $\approx 940$\,nm, as expected for the used CMs \cite{Silva2014a}. Overall, the standard deviation is larger for SD than for SHG d-scan, which we attribute to the lower sensitivity of the SD process to the spectral phase.

%Eventually, one loses information about the group delay difference between the two parts of the pulse, i.e., the well-compressed part (the main pulse) and the poorly-compressed wavelengths (those with diverging phase), which makes retrieval harder for those wavelengths.

%The phases show a good overall agreement, within their standard deviations, over the pulse spectrum ($\approx 450-1050$\,nm).
%The standard deviation is larger for SD d-scan, which we attribute to the different phase sensitivity of 3rd-order processes compared to 2nd-order ones.
%, hence no significant phase alterations appear to be taking place %either due to linear (dispersion) or nonlinear processes 
%in the 30\,{$\mu$}m slide.
%The reduction in the SD signal at longer wavelengths, possibly due to chromatic coupling of the SD signal into the spectrometer, results in an increased phase error in that region.
%The fast phase variation seen below $500$\,nm is due to the known divergence in the dispersion of the CMs. This makes retrieval harder for SHG and SD d-scan, which together with their different sensitivity to the incident pulse phase can result in the observed discrepancy for shorter wavelengths.
% \cite{Silva2014a}.

\begin{figure}[h]
\centering
\includegraphics[width=1.0\linewidth]{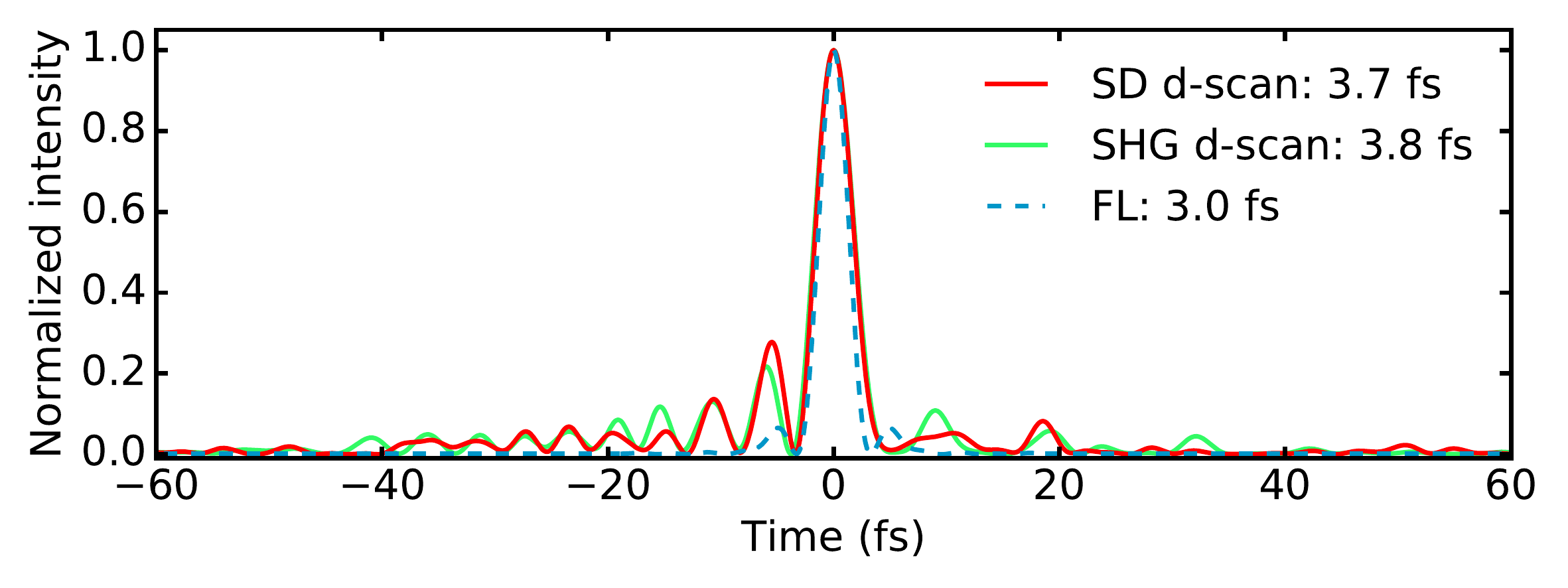}
\caption{ Retrieved temporal intensities for SD and SHG d-scan. The Fourier-limit (FL) of the CM-compressible portion of the spectrum ($\approx 500-1050$\,nm) is also shown for reference.}
\label{fig:intensity_plots}
\end{figure}

In the time domain (Fig.\,\ref{fig:intensity_plots}), the pulse intensity profiles retrieved by both methods are also very similar, including the pre- and post-pulse structure around the main pulse. The full-width at half-maximum (FWHM) pulse duration was found to be $3.7\pm0.3$\,fs for SD and $3.8\pm0.1$\,fs for SHG d-scan, in agreement with the value of $3.8\pm0.1$\,fs previously reported for the same HCF compressor before residual TOD compensation \cite{Silva2014a}.
%Propagation of these pulses through an adequate transparent medium, e.g. water \cite{Silva2014a} or z-cut KDP \cite{Miranda2017}, directly results in near-transform-limited, high-quality single-cycle 3\,fs pulses.

As mentioned previously, an advantage of the d-scan method is that no a priori knowledge of the spectral response function $R(\omega)$ is necessary for pulse retrieval, since the d-scan algorithm also retrieves it. Nonetheless, we derived a theoretical expression for $R(\omega)$, both for comparing with the retrieved response and to numerically explore the measurement of even broader bandwidth pulses using SD d-scan.
%in spite of the limited spectral range of the SD signal.
In Appendix A we show that the spectral response of the SD process is given by
\begin{equation}
R(\omega)=\frac{\omega^{2}}{n(\omega)}\left[n^{2}(\omega)-1\right]^{8}\mathrm{sinc}^{2}\left[\frac{\Delta k_{z}(\omega) L}{2}\right],
\label{eq:response_function}
\end{equation}
with $\Delta k_{z}$ the phase-mismatch along the propagation direction and $L$ the medium thickness. The mismatch can be written as $\Delta k_{z}(\omega)\approx \theta^{2}k(\omega)$, with $\theta \approx \theta_{\mathrm{ext}}/n$ the internal crossing angle in the medium and $\theta_{\mathrm{ext}}$ the external angle.
Since the SD d-scan trace was measured with the same intensity-calibrated spectrometer used for the linear spectrum, the retrieved response can be directly compared with that of the SD process alone. Figure\,\ref{fig:theoretical curves}a shows the retrieved SD d-scan response and the theoretical efficiency of SD in fused silica, calculated from \eqref{eq:response_function} for $L=30$\,{$\mu$}m and $\theta_{\mathrm{ext}}=1.4^{\circ}$. We see that the general trend of the experimental response roughly follows the theoretical prediction.
%The observed deviation is compatible with the mentioned chromatic coupling into the spectrometer. % which is not taken into account in the theoretical model.
%Nevertheless, \eqref{eq:response_function} shows how the noncollinear geometry simply changes the response function, which in d-scan does not translate into time smearing since d-scan traces are measured as a function of dispersion, not temporal delay.

\begin{figure}[h]
\centering
\includegraphics[width=1.0\linewidth]{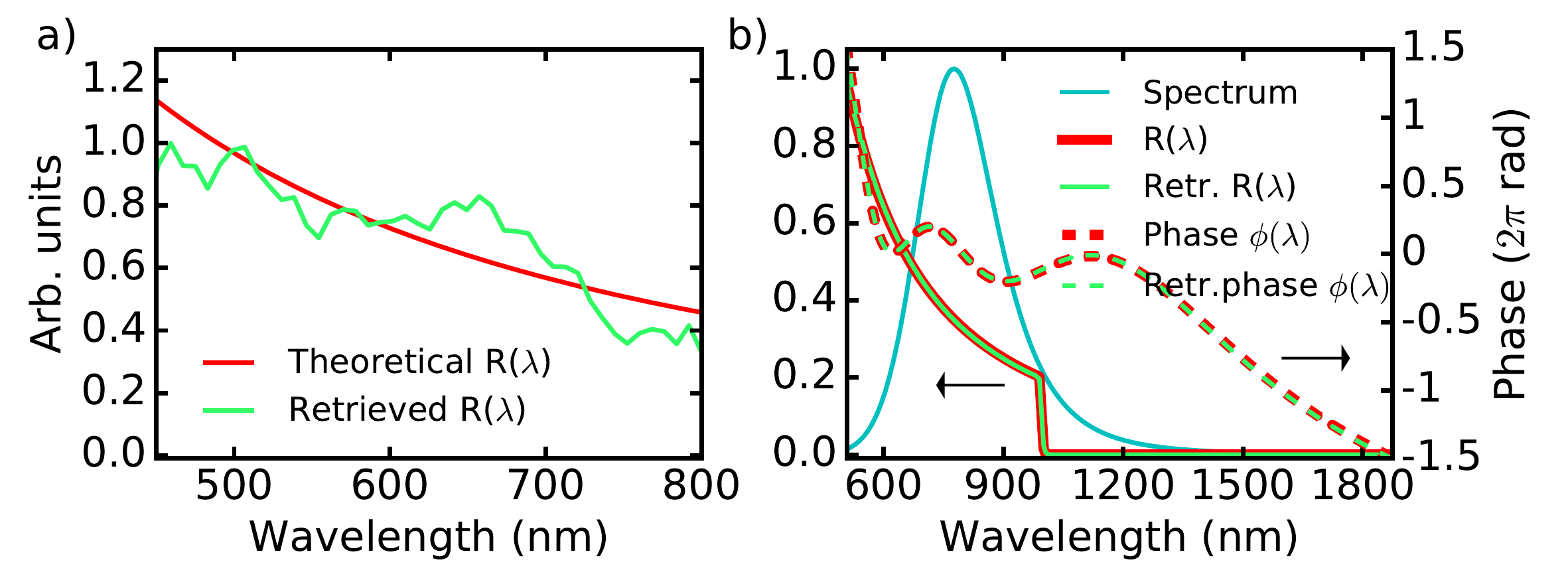}
\caption{SD d-scan: a) Theoretical and measured spectral response functions. b) Simulated retrieval over nearly 3 octaves.}
\label{fig:theoretical curves}
\end{figure}

To illustrate the possibility of multi-octave pulse measurement with SD d-scan, we simulated a sech2 spectrum centered at 800\,nm with a Fourier-limit of 2.7\,fs and a phase consisting of TOD and some ringing (see Fig.\,\ref{fig:theoretical curves}b). %The spectral wings of this pulse extend from 500\,nm to 1900\,nm.
The corresponding SD d-scan trace was calculated from \eqref{eq:SHG_d-scan} with the signal of \eqref{eq:SD/TG_d-scan} and the response function of \eqref{eq:response_function}, assuming a 30\,{$\mu$}m fused silica slide and a crossing angle $\theta_{\mathrm{ext}}=1.4^{\circ}$. The spectral response was clipped to zero after 1000\,nm to model the limited sensitivity of a Silicon detector, and a random noise baseline at -40\,dB was also added.
In these conditions, the d-scan algorithm successfully retrieved the spectral phase and the response function from 500 to 1900\,nm, i.e., over nearly 3 octaves (Fig.\,\ref{fig:theoretical curves}b).

\section{Conclusions}

In conclusion, we have developed a new d-scan technique for temporal pulse characterization based on self-diffraction (SD d-scan) and demonstrated it with over octave-spanning sub-4-fs pulses. The frequency-dependent efficiency of the SD process was derived theoretically and compared with the spectral response retrieved by the d-scan algorithm. SD d-scan has a robust inline implementation, is single-cycle and multi-octave capable, and should enable measuring ultra-broadband pulses with arbitrary polarization and in difficult spectral ranges, such as the ultraviolet and the deep-ultraviolet.
\\
\newline
%\section*{Funding}
\noindent \textbf{Funding.}
We acknowledge funding from UID/NAN/50024/2013 (FCT, Portugal) and FIS2013-41709P (MINECO, Spain).\\

%\section*{Acknowledgments}
\noindent \textbf{Acknowledgments.}
We thank Adam Wyatt for a brief but fruitful discussion on pulse propagation in thin nonlinear media.

\section*{Appendix A}

Analytical models of the spectral response of nonlinear processes are usually obtained assuming non-depletion of the input electric field and an estimated phase-mismatch \cite{trebino2012frequency}. Our derivation follows a similar treatment, with the necessary modifications, including the frequency dependence of the third-order nonlinear susceptibility. 
 %In the non-depletion regime, the power of the generated nonlinear signal is a small fraction of the input beam power.
Self-diffraction can be viewed as an interaction between two photons at $\omega_{1}$ and one photon at $\omega_{2}$, resulting in a frequency $\Omega = 2\omega_{1}-\omega_{2}$. When pertinent in this derivation, instead of considering the pair $(\omega_{1}, \omega_{2})$ as independent variables, we will consider the pair $(\Omega, \omega_{1})$.

The third-order susceptibility associated to SD, $\chi^{(3)}$, can be expressed as a product of linear susceptibilities, $\chi^{(1)}$, i.e. \cite{boyd2008nonlinear}
\begin{equation}
\chi^{(3)}(\Omega,\omega_{1},\omega_{1},-\omega_{2})\propto\chi^{(1)}(\Omega)\left[\chi^{(1)}(\omega_{1})\right]^{2}\chi^{(1)}(-\omega_{2}),
\label{Susceptibility1}
\end{equation}
where $\chi^{(1)}(\Omega)=n^{2}(\Omega)-1$, $\chi^{(1)}(-\omega_{2})=\chi^{*(1)}(\omega_{2})$, and for lossless materials, $\chi^{*(1)}(\omega_{2})=\chi^{(1)}(\omega_{2})$. Thus
\begin{equation}
\chi^{(3)}(\Omega,\omega_{1},\omega_{1},-\omega_{2})\propto\chi^{(1)}(\Omega)\left[\chi^{(1)}(\omega_{1})\right]^{2}\chi^{(1)}(\omega_{2}).
\label{eq:susceptibility2}
\end{equation}
If $\omega_{1}$ and $\omega_{2}$ are frequencies near $\Omega$, \eqref{eq:susceptibility2} becomes
%
%\begin{equation}
%\begin{split}
%\chi^{(3)}(\Omega,\omega_{1},\omega_{1},-\omega_{2}) & \approx
%\chi^{(3)}(\Omega,\Omega,\Omega,\Omega) \\
%&\propto\left[\chi^{(1)}(\Omega)\right]^{4}=\left[n^{2}(\Omega)-1\right]^{4}.
%\end{split}
%\label{susceptibility2}
%\end{equation}
%
\begin{equation}
\chi^{(3)} \approx \chi^{(3)}(\Omega,\Omega,\Omega,-\Omega) \propto\left[\chi^{(1)}(\Omega)\right]^{4}=\left[n^{2}(\Omega)-1\right]^{4}.
\label{susceptibility2}
\end{equation}
The phase-mismatch along the z-axis, $\Delta k_{z}$, is given by
\begin{equation}
\Delta k_{z}(\Omega,\omega_{1})=2k_{1z}(\omega_{1})-k_{2z}(2\omega_{1}-\Omega)-k_{3z}(\Omega),
\label{phase_mismatch}
\end{equation}
where $k_{1z}$, $k_{2z}$ and $k_{3z}$ are the projections of the incident and SD wavevectors along $z$ (see Fig.\,\ref{fig:setup}b).
%We can approximate \eqref{phase_mismatch} by a function of a single variable $\Omega$ (the {\sl generated} frequency).
Expanding $k_{1z}$ and $k_{2z}$ in a Taylor series around $\Omega$ (the {\sl generated} frequency),
\begin{align}
k_{1z}(\omega_{1}) &\approx k_{1z}(\Omega)+ \left. \frac{\mathrm{d}k_{1z}}{\mathrm{d}\omega_{1}} \right|_{w_{1}=\Omega}\left(\omega_{1}-\Omega\right) \\
k_{2z}(2\omega_{1}-\Omega) &\approx  k_{2z}(\Omega)+2 \left. \frac{\mathrm{d}k_{2z}}{\mathrm{d}\omega_{1}} \right|_{w_{1}=\Omega} \left(\omega_{1}-\Omega\right),
\label{taylor_expansion}
\end{align}
and substituting into \eqref{phase_mismatch}, we see that the terms with derivatives cancel each other up to 2nd-order corrections.
We assume that the amplitudes of the wavevectors are the same, i.e. $|{\boldsymbol{k}}_{1}(\Omega)| = |{\boldsymbol{k}}_{2}(\Omega)|= k(\Omega)$, so their projection in the z-axis is $k(\Omega) \cos(\theta/2)$ (Fig.\,\ref{fig:setup}b). For small crossing angles, the projection angle of ${\boldsymbol{k}}_{3}$ along $z$ can be obtained by using the law of sines to relate $\theta$ with $\gamma$ (Fig.\,\ref{fig:setup}b), the law of cosines to obtain $|{\boldsymbol{k}}_{3}|=k(\Omega) [5-4\cos(\theta)]^{1/2} \approx k(\Omega)$, and the linear approximation of the sine function to find $\gamma \approx 2\theta$, so \eqref{phase_mismatch} becomes
% For small crossing angles $|{\boldsymbol{k}}_{3}|\approx k(\Omega)$ and t
%
\begin{equation}
\begin{split}
\Delta k_{z}(\Omega)&\approx k(\Omega) \left[\cos(\theta/2) - \cos(3 \theta/2) \right] \approx \theta^{2} k(\Omega).
\end{split}
\label{delta_k}
\end{equation}

The electric field of a generic third-order process, after propagating a distance $L$, is given by \cite{trebino2012frequency} (p.~280)
\begin{equation}
\tilde{E}(L,\Omega)=i\frac{c\mu_{0}\Omega}{2n(\Omega)}\int_{0}^{^{L}} \tilde{P}^{(3)}(z,\Omega)e^{-ik_{3z}(\Omega)z}\mathrm{d}z,
\label{eq:electric field}
\end{equation}
where $\mu_{0}$ is the vacuum permeability and $\tilde{P}^{(3)}$ the nonlinear polarization. For SD, the polarization is given by
\begin{multline}
\tilde{P}^{(3)}(z,\Omega)=\\ \iint \chi^{(3)} \tilde{E}_1(z,\omega_{1}) \tilde{E}_{1}(z,\Omega-\omega_{1}+\omega_{2}) \tilde{E}_{2}^{*}(z,\omega_{2})\\
\times e^{i\left[k_{1z}(\omega_{1})+k_{1z}(\Omega-\omega_{1}+\omega_{2})-k_{2z}(\omega_{2}) \right] z} \mathrm{d}\omega_{1} \mathrm{d}\omega_{2},
\label{eq:polarization}
\end{multline}
where $\tilde{E}_{1,2}$ are the fields associated with $\boldsymbol{k}_{1,2}$ (see Fig. 1b).
Substituting \eqref{eq:polarization} into \eqref{eq:electric field} and integrating in $z$, we get
\begin{multline}
\tilde{E}(L,\Omega)=i\frac{c\mu_{0}\Omega L}{2n(\Omega)}\iint \chi^{(3)} \tilde{E}_{1}(\omega_{1}) \tilde{E}_{1}(\Omega-\omega_{1}+\omega_{2}) \\
\times \tilde{E}_{2}^{*}(\omega_{2})\mathrm{sinc}\left(\frac{\Delta k_{z} L}{2}\right) e^{i \Delta k_{z} L/2} \mathrm{d}\omega_{1} \mathrm{d}\omega_{2}.
\label{eq:electric_field2}
\end{multline}
The spectral intensity of this signal is $S_{\mathrm{meas}}\propto n \left|\tilde{E}\right|^{2}$.
%
%\begin{equation}
%\mathbb{I}_{meas}=\epsilon_{0}\frac{n_{4}(\Omega)Q(\Omega)}{c}\left|\tilde{E}_{4}\right|
%I_{\mathrm{meas}}=\epsilon_{0}n(\Omega)/c \left|\tilde{E}\right|^{2}.
%\end{equation}
%
Since $\Delta k_{z}$ and $\chi^{(3)}$ can be approximated by functions of one variable, $\Omega$, which plays no role in the integral of \eqref{eq:electric_field2}, the terms with these quantities can be factored out of the integral. This gives %By doing so (and discarding unnecessary constants, as we only need to establish proportionality between the relevant quantities) we arrive at
\begin{equation}
S_{\mathrm{meas}}\propto\frac{\Omega^{2}}{n(\Omega)}\left[n^{2}(\Omega)-1\right]^{8}\mathrm{sinc}^{2}\left[\frac{\Delta k_{z}(\Omega) L}{2}\right] \times S_{\mathrm{ideal}},
\label{eq:theoretical_intensity}
\end{equation}
with the ideal nonlinear SD spectral intensity, $S_{\mathrm{ideal}}$, defined as
\begin{equation}
S_{\mathrm{ideal}}\propto\left| \iint \tilde{E}_{1}(\omega_{1}) \tilde{E}_{1}(\Omega-\omega_{1}+\omega_{2}) \tilde{E}_{2}^{*}(\omega_{2}) \mathrm{d}\omega_{1} \mathrm{d}\omega_{2} \right|^{2},
\end{equation}
%and first term in \eqref{eq:theoretical_intensity} the response function of \eqref{eq:response_function}.
whereby we identify in \eqref{eq:theoretical_intensity} the spectral response of \eqref{eq:response_function}.

\end{document}